\documentclass[letterpaper,journal]{IEEEtran}

\usepackage{amsmath}    % need for subequations
\usepackage{amssymb}    % need for subequations
\usepackage{amsthm}     % need for subequations
\usepackage{graphicx}   % need for figures
\usepackage{verbatim}   % useful for program listings
\usepackage{subfigure}  % use for side-by-side figures
\usepackage{epstopdf}
\usepackage[utf8]{inputenc}
\usepackage{stfloats}
\usepackage{bigstrut}
\usepackage{cite}

\begin{document}
\title{Polar Codes for Nonasymmetric Slepian--Wolf Coding}

\author{\IEEEauthorblockN{Saygun \"Onay} \\
\IEEEauthorblockA{Bilkent University, Ankara, Turkey}%
\thanks{This work was supported by The Scientific and Technological Research
Council of Turkey (T\"UB\.ITAK) under Project 110E243.}}

\maketitle

\begin{abstract}

A method to construct nonasymmetric distributed source coding (DSC) scheme using polar codes which can achieve any point on the dominant face of the Slepian--Wolf (SW) rate region for sources with \emph{uniform} marginals is considered. In addition to nonasymmetric case, we also discuss and show explicitly how asymmetric and single source compression is done using successive cancellation (SC) polar decoder. We then present simulation results that exhibit the performance of the considered methods.
\end{abstract}


\IEEEpeerreviewmaketitle

\section{Introduction}\label{sec:introduction}

In this paper we present a method to achieve any point on the ``dominant" face of the Slepian--Wolf (SW) achievable rate region using polar codes for the case of sources with \emph{uniform} marginals. SW coding \cite{slepian_noiseless_1973} refers to the distributed compression of a memoryless source pair $(X,Y)$  $\sim p_{X,Y}(x,y)$. The problem setting assumes separate encoding but joint decoding of sources $X$ and $Y$. The surprising result of the SW theorem states that a total rate $H(X,Y)$ is sufficient even with separate encoding. In this work we assume $p_{X,Y}$ is a distribution on ${\cal X} \times {\cal X}$ where ${\cal X} = \{0,1\}$. Let $(x^N, y^N) \in {\cal X}^N \times {\cal X}^N$ be a pair of $N$-vectors obtained by repeated independent drawings of source pair $(X,Y)$. Then, $X$-encoder performs the mapping $f_X : {\cal X}^N \rightarrow \{1,2, \ldots, 2^{NR_X} \}$ and $Y$-encoder performs the mapping $f_Y : {\cal X}^N \rightarrow \{1,2, \ldots, 2^{NR_Y} \}$. The decoder observing both of the sequences performs the mapping $g : \{1,2, \ldots, 2^{NR_X} \} \times \{1,2, \ldots, 2^{NR_Y} \} \rightarrow {\cal X}^N \times {\cal X}^N$. As $N \rightarrow \infty$, such mappings with vanishingly small decoding error probability exist if rate pair $(R_X, R_Y)$ is inside the \emph{achievable rate region} (SW rate region) described by the inequalities $R_X \ge H(X|Y)$, $R_Y \ge H(Y|X)$ and $R_X+R_Y \ge H(X,Y)$. The corner points $(R_X,R_Y) = (H(X|Y),H(Y))$ and $(R_X,R_Y) = (H(X),H(Y|X))$ on the achievable rate boundary are also referred to as ``asymmetric" operating points. And any point on the line segment between these corner points (``dominant" face), where $R_X+R_Y = H(X,Y)$, is also referred to as a ``nonasymmetric" operating point.

In the past decade, starting with the pioneering work of Pradhan and Ramchandran \cite{pradhan_distributed_1999}, an extensive literature on applying channel codes for practical implementation of SW coding has been developed. Schemes utilizing turbo and LDPC codes in both asymmetric  \cite{liveris_compression_2002-1, liveris_distributed_2003} and nonasymmetric \cite{schonberg_distributed_2004, stankovic_code_2006} SW problems were constructed. Also, ``flexible rate" schemes using LDPC \cite{varodayan_rate-adaptive_2005} and turbo \cite{roumy_rateadaptive_2007, zamani_flexible_2009} codes were devised. A flexible rate code refers to a SW coding scheme which has means to vary its total rate without much performance loss. The nonasymmetric schemes in \cite{schonberg_distributed_2004, stankovic_code_2006} are based on ``channel code partitioning" idea introduced in \cite{pradhan_distributed_2000}. A single channel code generator matrix is partitioned into two to be used by $X$ and $Y$ encoders. The partition is done in such a way that the desired rate allocation is achieved. Another method to achieve nonasymmetric rate allocation was introduced in \cite{gehrig_symmetric_2005}. It uses the same channel code for both encoders. It is the method used in this paper and its details are given in Section \ref{sec:nsw}.

Polar coding \cite{arikan_channel_2009}, recently discovered by Arıkan, is the first provably capacity--achieving coding method with low encoding and decoding complexity for the class of binary--input discrete memoryless channels. Shortly after its discovery, a number of work has been published which showed that polar codes are also provably optimal for source coding, \emph{asymmetric} Slepian--Wolf and Wyner--Ziv problems \cite{korada_polar_2009, korada_polar_2010-1, hussami_performance_2009, cronie_lossless_2010, arikan_source_2010}. Our main contribution in this work is to devise a simple practical \emph{nonasymmetric} SW scheme using polar codes for the case of sources with \emph{uniform} marginals by utilizing the framework of \cite{gehrig_symmetric_2005}.

The paper is organized as follows. First, we briefly discuss how to use polar encoder / SC decoder pair in practical asymmetric SW and single source compression settings. Then, we proceed to describing how to apply the framework of \cite{gehrig_symmetric_2005} on polar codes to achieve nonasymmetric SW coding of \emph{uniform} sources. Lastly, we present simulation results that exhibit the performance of the considered schemes. We use SC list (SCL) decoder \cite{tal_list_2011} with CRC to achieve the best possible results. The method of adding a simple CRC to information bits and using the SCL decoder in conjunction with this CRC was proposed by the authors of \cite{tal_list_2011}. The advantage of this method is that it improves the performance for short to moderate block lengths with almost no extra complexity. We assume that the reader is familiar with polar codes, especially the concepts in \cite{arikan_channel_2009, arikan_source_2010}. The notations used for vectors and matrices are similar to those in \cite{arikan_channel_2009} and \cite{arikan_systematic_2011}.

\section{Asymmetric Slepian--Wolf and Single Source Compression}\label{sec:ssc_asw}

In this work, the marginals of both sources $X$ and $Y$ are assumed to be uniform. Let $Y=X\oplus Z$, where $Z \sim$ Ber$(p)$. In other words, $Y$ is the corrupted version of $X$ by a BSC($p$). Thus, $H(X|Y)=H(Y|X)={\cal H}(p)$, where ${\cal H}(p)$ is the binary entropy function $-p\cdot \log_2(p)-(1-p)\cdot \log_2(1-p)$. For asymmetric SW setting, side information (SI) $y^N$ is assumed to be present at the decoder error--free. Then, this setting is actually not much different than a channel decoding problem. Given the observation $y^N$, which is a corrupted version of $x^N$ by a BSC($p$), it is nothing but a channel decoding problem to recover $x^N$. The only difference compared to a usual channel decoding is that the coset in which the search to be performed is not the zero syndrome one but given by the syndrome of $x^N$. Thus, the compression operation is nothing but calculating the syndrome $s^{N-K}$ of $x^N$. One major advantage of polar codes is that this nonstandard decoding is readily implemented with a SC decoder \cite{arikan_channel_2009}.

\begin{figure}[t]
	\centering
	\subfigure[Source Encoder]{
		\label{fig:single:sub1}\includegraphics[width=3.2in]{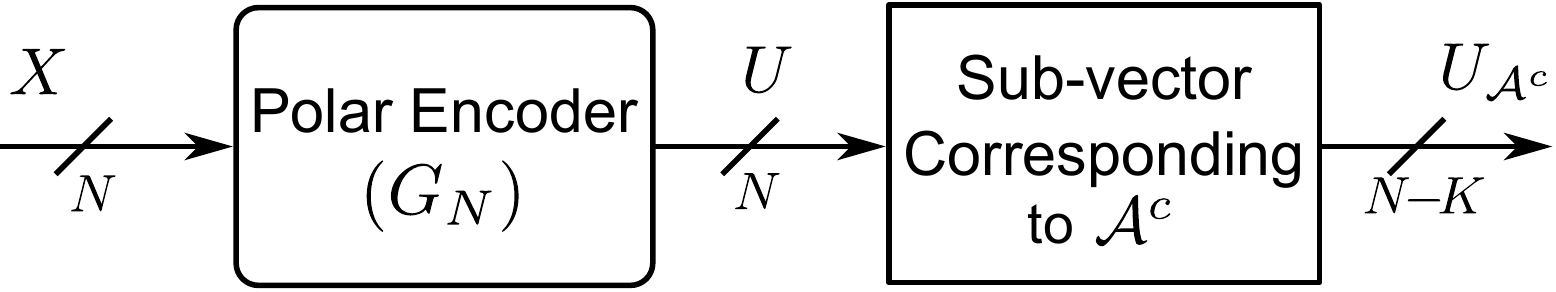}
	}\\
	\subfigure[Source Decoder]{
		\label{fig:single:sub2}\includegraphics[width=3.2in]{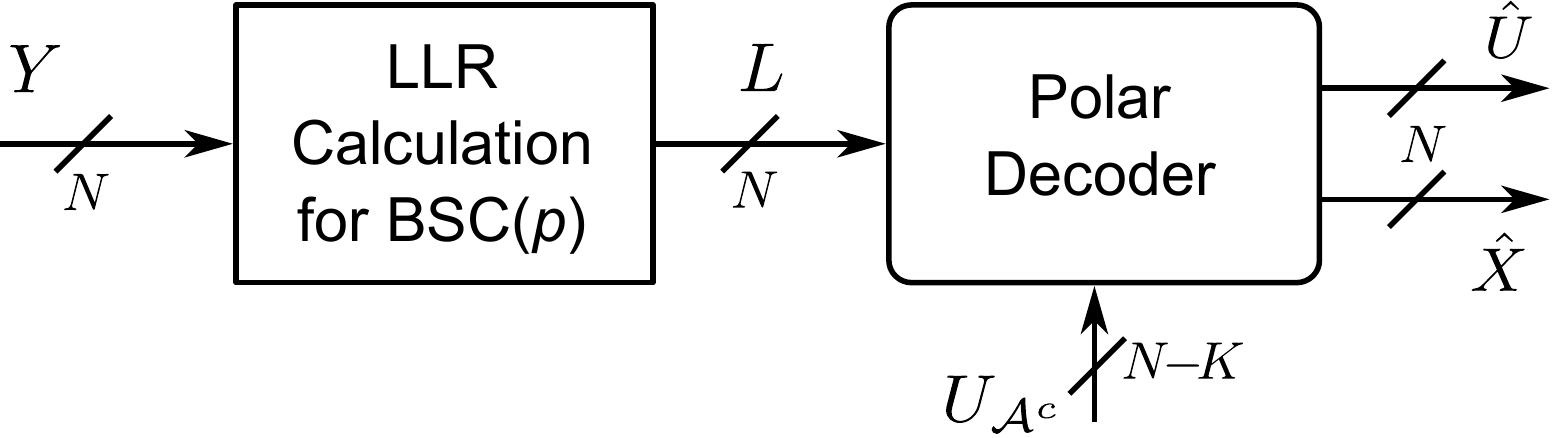}
	}
	\caption{Polar codes for asymmetric SW compression.}
	\label{fig:single}
	\vspace{-2mm}
\end{figure}

A polar code is identified by a parameter set $(N,K,{\cal A}, u_{A^c})$, where $N$ is the block length, $K$ is the code dimension, ${\cal A}$ is the information set of size $K$ and $u_{A^c}$ is the frozen bits vector of size $N-K$. The frozen bits $u_{{\cal A}^c}$ identify a coset of the linear block code with generator matrix $(G_N)_{\cal A}$ \cite{arikan_channel_2009}. $(G_N)_{\cal A}$ denotes the submatrix of $G_N$ formed by the rows with indices in ${\cal A}$. Therefore, a parity check matrix can be found so that $u_{{\cal A}^c}$ is the syndrome of this code. Since $G_N^{-1}=G_N$, the syndrome calculation is just a polar encoding operation followed by extracting the bits corresponding to the frozen indices:
\begin{align}\label{eqn:synd}
s^{N-K}=u_{{\cal A}^c}=(x^NG^{-1}_N)_{{\cal A}^c}=(x^NG_N)_{{\cal A}^c}.
\end{align}

The encoding and decoding operations are summarized in \figurename \ref{fig:single}. Encoding realizes the operation defined in \eqref{eqn:synd}. Decoding is done using the SCL decoder which outputs estimates of both $\mathbf{x}$ and $\mathbf{u}$. $\hat{\mathbf{x}}$ is the estimate of the uncompressed bit sequence. ``LLR calculation" block calculates the LLRs according to the \emph{assumed} correlation between sources $X$ and $Y$, which is a BSC($p$):
\begin{align}\label{eqn:llr_zero}
L_i = \begin{cases} (-1)^{y_i} \cdot \log \frac{1-p}{p}, &\: i \in \{1,\ldots,N-l_{crc}\} \\
													  0, &\: i \in \{N-l_{crc}+1,\ldots,N\},
\end{cases}
\end{align}
where $l_{crc}$ denotes the length of CRC bits, if any. CRC is used in conjuction with SCL decoder to increase performance. If it is used, the length of information block is reduced to $N'=N-l_{crc}$. Note that in \eqref{eqn:llr_zero}, while the statistics of the first $(N-l_{crc})$ bits are known $\left(\text{Ber}(p)\right)$ and used for decoding, the statistics of the CRC bits are assumed to be uniform. 

For single source compression, the SI vector $y^N$ is replaced with all--zeros vector $0^N$ at the input of the decoder.

\section{Nonasymmetric Slepian--Wolf Compression}\label{sec:nsw}

A method to construct nonasymmetric SW scheme from a single channel code for the case of \emph{uniformly} distributed sources using syndrome approach was proposed in \cite{gehrig_symmetric_2005}. This method was recently used in \cite{zamani_flexible_2009} to construct a both nonasymmetric and rate--compatible SW scheme using turbo codes. In this section, we apply the method of \cite{gehrig_symmetric_2005} to construct a nonasymmetric SW setting using polar codes.

\begin{figure}[h]
	%\vspace{-4mm}
	\centering
	\includegraphics[width=2.2in]{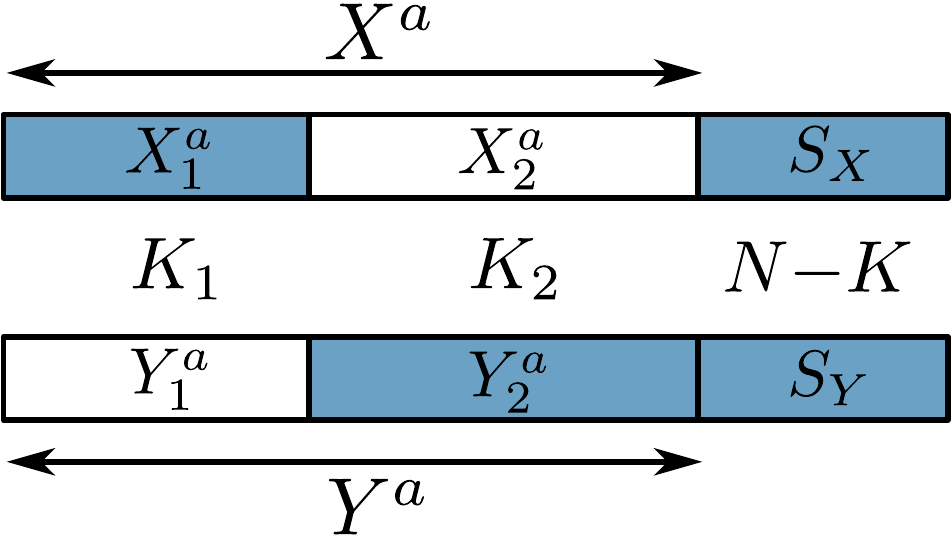}
	\caption{Encoding for nonasymmetric SW with method of \cite{gehrig_symmetric_2005}.}
	\label{fig:nonasym1}
	\vspace{-2mm}
\end{figure}

The method of \cite{gehrig_symmetric_2005} can be summarized as follows. Consider two uniform i.i.d. distributed and correlated $N$--vectors $\mathbf{x}=[\mathbf{x}^a\quad \mathbf{x}^b]$ and $\mathbf{y}=[\mathbf{y}^a\quad \mathbf{y}^b]$, where $\mathbf{x}^a$ represents the first $K$ bits and $\mathbf{x}^b$ represents the last $N-K$ bits of vector $\mathbf{x}$ (the same applies to $\mathbf{y}$). Also, let $\mathbf{x}^a = [\mathbf{x}^a_1\quad \mathbf{x}^a_2]$ and $\mathbf{y}^a = [\mathbf{y}^a_1\quad \mathbf{y}^a_2]$.  $\mathbf{x}^a_1$ represents the first $K_1$ bits and $\mathbf{x}^a_2$ represents the last $K_2$ bits of $\mathbf{x}^a$ (the same applies to $\mathbf{y}^a$), where $K_1+K_2=K$. Let $G$ be $K\times N$ generator matrix and $H=[H_a\quad H_b]$ be $(N-K)\times N$ parity check matrix of the channel code. Also assume that $H_a$ is $(N-K)\times K$ submatrix and $H_b$ is $(N-K)\times (N-K)$ submatrix of $H$ such that $H_b$ is selected as nonsingular. Notice that the systematic version of a code is a special case with $H_b = I_{N-K}$.
The syndromes of $\mathbf{x}$ and $\mathbf{y}$ are calculated as $\mathbf{s_x} = \mathbf{x} H^T = \mathbf{x}^a H_a^T\oplus \mathbf{x}^b H_b^T$ and $\mathbf{s_y} = \mathbf{y} H^T = \mathbf{y}^a H_a^T\oplus \mathbf{y}^b H_b^T$, respectively. Then, $X$--encoder sends $(\mathbf{x}^a_1,\mathbf{s_x})$ and $Y$--encoder sends $(\mathbf{y}^a_2,\mathbf{s_y})$. The information sent by both encoders are marked with the shaded regions in \figurename \ref{fig:nonasym1}. By varying $K_1$ (keeping $K_1 + K_2 = K$ at constant) different points on the SW rate region can be reached. Since $X$ and $Y$ are assumed to be uniform i.i.d. sources, $H(X)=H(Y)=1$. Then, the code is adjusted so that $(N-K) \geq N H(X|Y)$. Thus, the total rate stays above entropy bound:
\begin{align*}
NR&=N[R_X+R_Y]=(K_1+(N-K))+(K_2+(N-K)) \\
NR&=N + (N-K) \geq N[H(Y) + H(X|Y)] = N H(X,Y).
\end{align*}

\begin{figure}[t]
	\centering
	\includegraphics[width=3.2in]{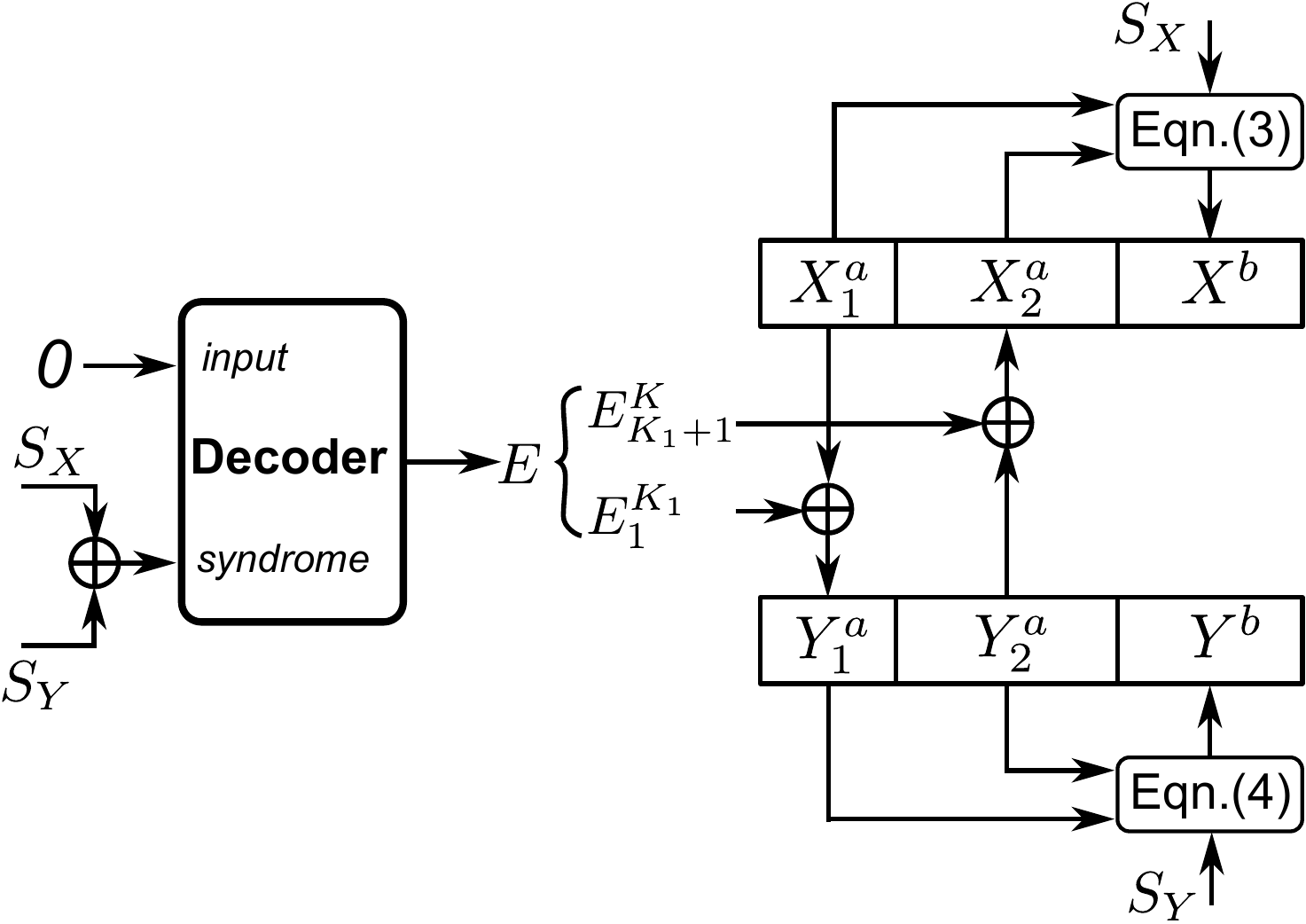}
	\caption{Decoding for nonasymmetric SW with method of \cite{gehrig_symmetric_2005}.}
	\label{fig:nonasym2}
	\vspace{-4mm}
\end{figure}

The decoding of the above scheme, which is depicted in \figurename \ref{fig:nonasym2}, is done as follows. Let $\mathbf{e} = \mathbf{x} \oplus \mathbf{y}$ be the error vector. Then, $\mathbf{s_e} = \mathbf{e} H^T = (\mathbf{x}\oplus \mathbf{y}) H^T = \mathbf{s_x}\oplus \mathbf{s_y}$. The channel decoder is supplied with all--zeros vector as input and $\mathbf{s_e}$ as the coset index. The estimate $\hat{\mathbf{e}}$ is obtained as the output. With this estimated error pattern, $\mathbf{x}^a_2$ and $\mathbf{y}^a_1$ can be recovered using $\mathbf{y}^a_2$ and $\mathbf{x}^a_1$, respectively. Finally, $\mathbf{x}^b$ and $\mathbf{y}^b$ are obtained as 
\begin{align}
\mathbf{x}^b &= (\mathbf{s_x}\oplus \mathbf{x}^a H_a^T) (H_b^T)^{-1}, \\
\mathbf{y}^b &= (\mathbf{s_y}\oplus \mathbf{y}^a H_a^T) (H_b^T)^{-1}.
\end{align}
Note that, although it is not shown explicitly in \figurename \ref{fig:nonasym2}, likelihood calculation of the the all--zeros vector input to the decoder is done using the crossover parameter $p$ of the virtual BSC between sources $X$ and $Y$ as given in \eqref{eqn:llr_zero}.

Now, this method cannot be applied to the standard form of polar codes in \cite{arikan_channel_2009}. However, the systematic version of polar codes \cite{arikan_systematic_2011} which was introduced recently can be used. The systematic polar coding is defined as follows. The codeword $\mathbf{x}$ is split into two parts $\mathbf{x} = (\mathbf{x}_{\cal B}, \mathbf{x}_{{\cal B}^c})$, where ${\cal B}$ is a subset of $\{1,\ldots,N\}$ such that $|{\cal B}|=|{\cal A}|$ and submatrix $G_{{\cal A}{\cal B}}$ is invertible. $G_{{\cal A}{\cal B}}$ consists of the array of elements $(G_{i,j})$ with $i \in {\cal A}$ and $j \in {\cal B}$. Then, the ``systematic" and ``parity" part of the codeword $\mathbf{x}$ can be written as
\begin{align}
\mathbf{x}_{\cal B} &= \mathbf{u}_{\cal A}G_{{\cal A}{\cal B}}+\mathbf{u}_{{\cal A}^c}G_{{\cal A}^c{\cal B}} \label{eqn:x_b}\\
\mathbf{x}_{{\cal B}^c} &= \mathbf{u}_{\cal A}G_{{\cal A}{\cal B}^c}+\mathbf{u}_{{\cal A}^c}G_{{\cal A}^c{\cal B}^c},
\label{eqn:x_bc}
\end{align}
respectively. The difference of this definition compared to a usual systematic code definition is that the systematic bits do not constitute the first $K$ bits of the codeword $\mathbf{x}$ but rather a different subset of locations identified by the index set ${\cal B}$. Now with the above definitions, a systematic encoder defined with parameter $({\cal B}, \mathbf{u}_{{\cal A}^c})$ can implement the mapping $\mathbf{x}_{\cal B} \mapsto \mathbf{x} = (\mathbf{x}_{\cal B}, \mathbf{x}_{{\cal B}^c})$ by computing
\begin{align}
\mathbf{u}_{\cal A} = (\mathbf{x}_{\cal B} \oplus \mathbf{u}_{{\cal A}^c} G_{{\cal A}^c {\cal B}}) (G_{{\cal A}{\cal B}})^{-1}
\end{align}
and then inserting $\mathbf{u}_{\cal A}$ into \eqref{eqn:x_bc} to obtain $\mathbf{x}_{{\cal B}^c}$.

Now returning back to the nonasymmetric SW method of \cite{gehrig_symmetric_2005} described above, we set $\mathbf{x}^a = \mathbf{x}_{\cal B}$, $\mathbf{x}^b = \mathbf{x}_{{\cal B}^c}$ and $\mathbf{s_x} = \mathbf{u}_{{\cal A}^c}$. This way we fulfill the requirements of the method such that when $\mathbf{x}^a$ is decoded using the estimated error vector $\hat{\mathbf{e}}$, the rest, $\mathbf{x}^b$, can be recovered from $\mathbf{x}^a$ and $\mathbf{s_x}$. Given $\mathbf{x}^a=\mathbf{x}_{\cal B}$ and $\mathbf{s_x} = \mathbf{u}_{{\cal A}^c}$, computing $\mathbf{x}^b = \mathbf{x}_{{\cal B}^c}$ and $\mathbf{u}_{\cal A}$ is nothing but a systematic polar encoding operation summarized above. And it can be done efficiently using a SC polar decoder \cite{arikan_systematic_2011}. For the standard form of polar codes defined in \cite{arikan_channel_2009}, the index set ${\cal B}$ can be selected as the permuted version of ${\cal A}$. This permutation corresponds to the \textit{bit-reversal} operation.

The use of CRC with this nonasymmetric scheme is also possible. For both sources, $N'=N-l_{crc}$ length information blocks are completed to $N$ with $l_{crc}$ bits of CRC. Since CRC operation is linear, the CRC of error vector $\mathbf{e} = \mathbf{x} \oplus \mathbf{y}$ must also check. Thus, the SCL channel decoder can use this information when estimating the error vector.

\section{Simulation Results}

In this section, we present simulation results on performances of the source coding methods discussed. The correlation model between sources $X$ and $Y$ is given as $Y = X \oplus Z$, where $Z \sim \text{Ber}(p)$. In all of the plots, the rates of codes are kept at a defined constant value while $p$ is varied to achieve different $H(X|Y)$ points. For all of the SW schemes the plotted BER corresponds to the averaged value over $X$ and $Y$ sources. The polar decoder used is the SCL decoder of \cite{tal_list_2011}. To improve the performance, a 16-bit CRC (CCITT) is added. The list decoder selects the output from the final list with the aid of CRC. Note that, for source coding, CRC is appended to the ``codeword" vector $\mathbf{x}$ as opposed to channel coding case where it is appended to ``information" vector $\mathbf{u}_{\cal A}$. The list size is set to 32 for all cases. The code construction is done via the method proposed in \cite{tal_how_2011} and optimized to $p=0.09$ for $R=0.5$, $p=0.03$ for $R=0.25$ and $p=0.013$ for $R=0.125$.

\begin{figure}[t]
	\centering
	\includegraphics[width=3.2in]{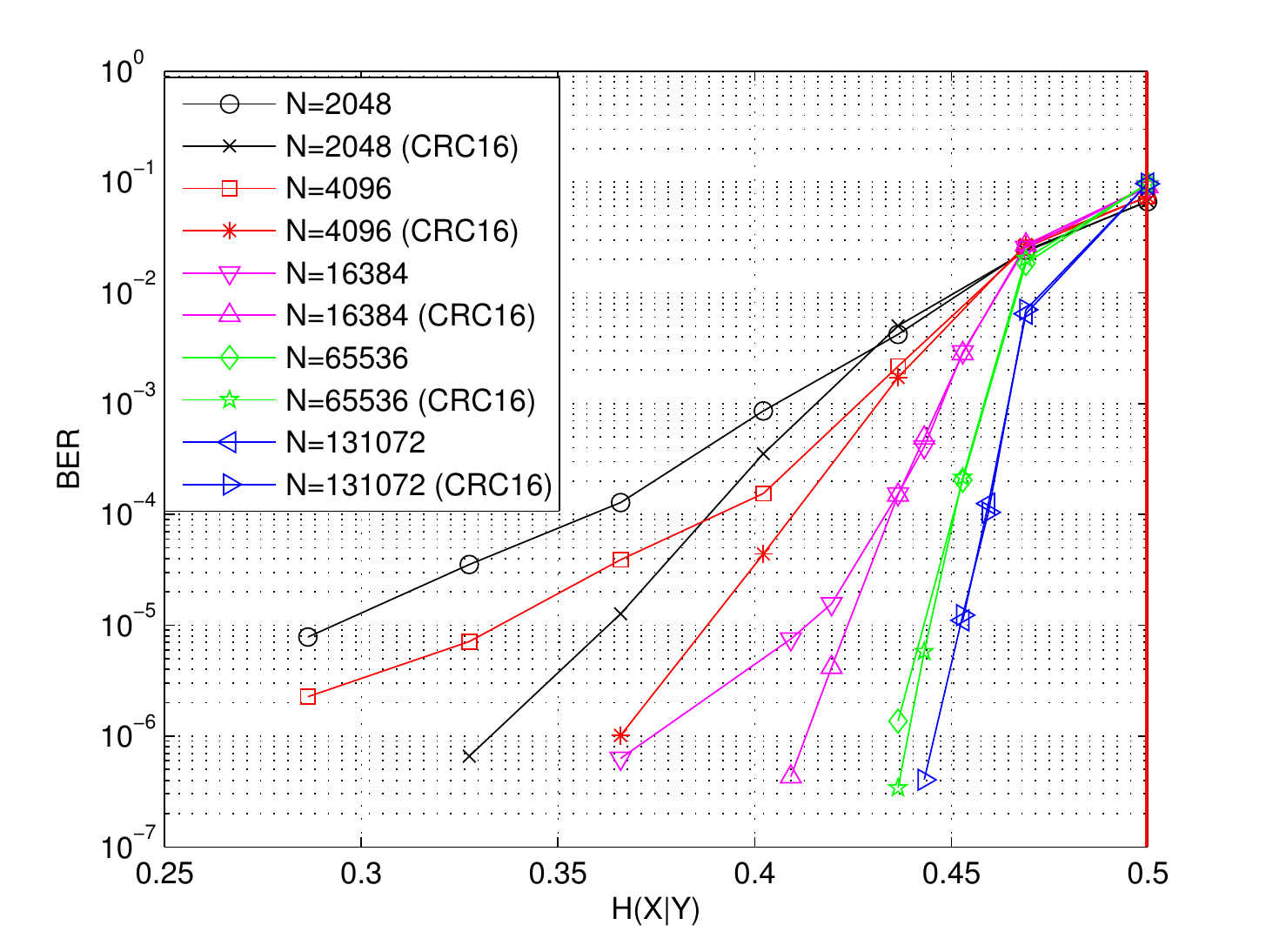}
	\caption{Asymmetric SW bit error rate $(R_X=0.5)$.}
	\label{fig:sim:asw}
\vspace{-4mm}
\end{figure}

\subsection{Asymmetric Slepian--Wolf and Single Source Compression}

As it is pointed out in Section \ref{sec:ssc_asw}, single source coding can be considered as a special case of asymmetric SW coding. Single source compression of an i.i.d. source $X$ with $H(X)={\cal H}(p)$ and asymmetric SW coding of i.i.d. pair $(X,Y)$ with uniform marginals and $H(X|Y)=H(Y|X)={\cal H}(p)$ are essentially the same. Therefore, their simulation performances come out to be identical, as expected. In \figurename \ref{fig:sim:asw}, BER curves of asymmetric SW setting for $R_X=0.5$ are shown. Results for different code block lengths and rate values are given in Table \ref{tab:asw}. The values in the table show the entropy of source correlation $({\cal H}(p)=H(X|Y))$ when a BER of $10^{-5}$ is achieved. For example, the gap to SW bound for $R_X=0.5$ and $N=65536$ is 0.056.

\begin{table}[h]
  \centering
  \caption{Asymmetrical SW performance $(H(X|Y) \: \text{values for a BER of } 10^{-5})$}
    \begin{tabular}{l|ccccc}
    $\mathbf{R_X}$ \textbackslash \: \textbf{N} & 2048  & 4096  & 16384 & 65536 & 131072 \bigstrut[b]\\
    \hline \hline
    $0.5$ & 0.360  & 0.386 & 0.423 & 0.444 & 0.453 \bigstrut[t]\\
    \hline
    $0.25$ & 0.118  & 0.153 & 0.190 & 0.205 & 0.210 \bigstrut[t]\\
    \hline
    $0.125$ & 0.032  & 0.044 & 0.075 & 0.096 & 0.101 \bigstrut[t]\\
    \end{tabular}%
  \label{tab:asw}%
  \vspace{-4mm}
\end{table}%

\subsection{Nonasymmetric Slepian--Wolf Compression}

\begin{figure}[t]
	\centering
	\includegraphics[width=3.2in]{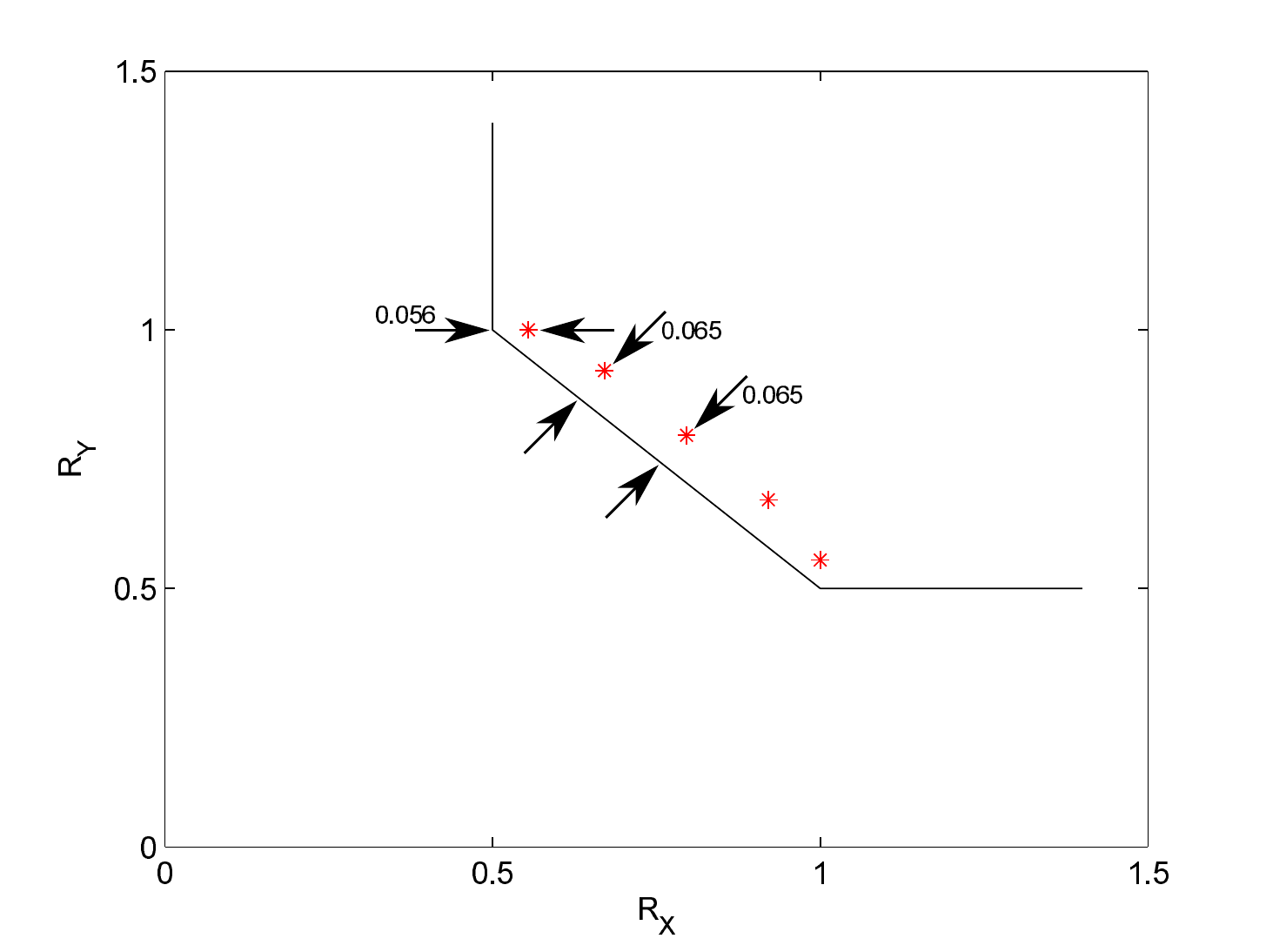}
	\caption{Nonasymmetric method for $N=65536$ together with the SW bound.}
	\label{fig:sim:nsw}
	\vspace{-4mm}
\end{figure}

The performance of nonasymmetric scheme for ${\cal H}(p)=0.5$ and $N=65536$ is presented in \figurename \ref{fig:sim:nsw}. Results for different block lengths are given in Table \ref{tab:nsw}. A BER of $10^{-5}$ is considered to be lossless when determining the rate points. The performance %of the nonasymmetric method of Section \ref{sec:nsw} with%
with rates allocated such that it results in asymmetric setting $(R_X,R_Y)=(0.5,1)$, is identical to the results of asymmetric method of Section \ref{sec:ssc_asw} given in Table \ref{tab:asw} (a gap of 0.056). This is expected, because when asymmetric rate allocation is made in the nonasymmetric method, $y^N$ is recovered perfectly. Also, only the syndrome vector is sent from the $X$-encoder. Thus, for asymmetric rate allocation, the method of Section \ref{sec:nsw} reduces to the method of Section \ref{sec:ssc_asw}. Two other operating points, corresponding to the symmetric rate $(0.75,0.75)$ and a nonasymmetric rate $(0.625,0.875)$, are also marked on \figurename \ref{fig:sim:nsw}. The performance of these rates are slightly inferior to the asymmetric case. This is expected, since as opposed to asymmetric case where no error is made for the source $Y$, in nonasymmetric cases estimation of $Y$ is also prone to errors, furthermore these errors propagate to the recovery of $X$. Also note that the performance is the same for all nonasymmetric points.

\section{Conclusion}

We started by reviewing asymmetric SW and single source compression using polar codes. Then, using the general framework of \cite{gehrig_symmetric_2005}, we showed how polar codes can be used for constructing nonasymmetric SW scheme for \emph{uniform} sources. A successive cancellation list decoder with addition of 16-bit CRC is used to achieve best performances. Although the performances are very good, they are slightly inferior to the best performances reported in the literature using turbo and LDPC codes \cite{liveris_compression_2002-1, stankovic_code_2006, zamani_flexible_2009}. However, there are some advantages to using polar codes. No modification is needed to the SC decoder since syndrome decoding is readily implemented. The length of the syndrome can be modified easily and incrementally, which gives rise to flexible rate adaptation. This might be an incentive to use polar codes in varying correlation conditions. Also, exploring advantages of polar codes against turbo and LDPC codes in terms complexity and latency remains as a future research topic.

\begin{table}[t]
  \centering
  \caption{Nonasymmetrical SW performance for $R=1.5$ $(H(X,Y) \: \text{values for a BER of } 10^{-5})$}
    \begin{tabular}{l|cccc}
    $\mathbf{(R_X,R_Y)}$ \textbackslash \: \textbf{N} & 2048  & 4096  & 16384 & 65536 \bigstrut[b]\\
    \hline \hline
    $(0.500,1.000)$ & 1.361  & 1.388 & 1.424 & 1.444 \bigstrut[t]\\
    \hline
    $(0.625,0.875)$ & 1.321  & 1.349 & 1.402 & 1.435 \bigstrut[t]\\
    \hline
    $(0.750,0.750)$ & 1.321  & 1.349 & 1.402 & 1.435 \bigstrut[t]\\
    \end{tabular}%
  \label{tab:nsw}%
  \vspace{-4mm}
\end{table}%

%\vfill

\bibliographystyle{ieeetr}

\end{document}